\documentclass[a4paper]{article}

\usepackage{INTERSPEECH2021}
\usepackage{multirow}
\usepackage{subfigure}
\usepackage{balance}
\usepackage{threeparttable}% add footnote within tables
\usepackage{multirow}
\usepackage{xcolor}
\usepackage{color}
\usepackage{soul}
\usepackage{etoolbox}

\newbool{showComments}
\booltrue{showComments}
\ifbool{showComments}{%

\newcommand{\jh}[1]{\sethlcolor{cyan}\hl{[JH: #1]}}
\newcommand{\tx}[1]{\sethlcolor{pink}\hl{[Tong: #1]}}
\newcommand{\cm}[1]{\sethlcolor{yellow}\hl{[Cecilia: #1]}}
\newcommand{\td}[1]{\sethlcolor{green}\hl{[Sally: #1]}}
\newcommand{\lo}[1]{\sethlcolor{orange}\hl{[Lorena: #1]}}
}
{

\newcommand{\jh}[1]{}
\newcommand{\tx}[1]{}
\newcommand{\cm}[1]{}
\newcommand{\td}[1]{}
\newcommand{\lo}[1]{}
}

\title{Uncertainty-Aware COVID-19 Detection from Imbalanced Sound Data}
\name{Tong Xia, Jing Han, Lorena Qendro, Ting Dang, Cecilia Mascolo}
\address{Department of Computer Science and Technology, University of Cambridge, UK}
\email{tx229@cam.ac.uk }

\begin{document}

\maketitle
\begin{abstract}
% In the era of the COVID-19 pandemic, researchers have made considerable efforts to develop fast and accurate testing tools, to speed up the early identification of COVIOD-19 patients. In particular, speech and other human generated audio signals have been explored and shown huge promise to achieve scalable and prompt digital pre-screening of the disease. However, there are still two unsolved issues hindering the practice. First, collected data are often imbalanced, with a considerably smaller proportion of positive samples than negative ones, making it harder to learn representative features for the minority class. Second, an AI prediction should provide not only a binary decision; in real life, especially for clinical applications, how certain the prediction is should also be taken into account. To tackle these, we propose an ensemble learning-based framework for sound-based COVID-19 detection. On one hand, multiple models are learnt from different but balanced subsets of the original imbalanced data, and better predictive performance can be obtained by combining these models than a single model. On the other hand, we compute an estimate of uncertainty from the disagreement across these models, and observed that false predictions often yield higher uncertainty. This enables us to suggest the users with certainty higher then a threshold to repeat audio test on their phones or to take clinical tests if digital diagnosis still fails after trying several times. This study paves the way for a more robust sound-based COVID-19 automated screening system. 

Recently, sound-based COVID-19 detection studies have shown great promise to achieve scalable and prompt digital pre-screening. However, there are still two unsolved issues hindering the practice.  First, collected datasets for model training are often imbalanced, with a considerably smaller proportion of users tested positive, making it harder to learn representative and robust features. Second, deep learning models are generally overconfident in their predictions. Clinically, false predictions aggravate healthcare costs. Estimation of the uncertainty of screening would aid this.
To handle these issues, we propose an ensemble framework where multiple deep learning models for sound-based COVID-19 detection are developed from different but balanced subsets from original data. As such, data are utilized more effectively compared to traditional up-sampling and down-sampling approaches: an AUC of 0.74 with a sensitivity of 0.68 and a specificity of 0.69 is achieved. Simultaneously, we estimate uncertainty from the disagreement across multiple models. It is shown that false predictions often yield higher uncertainty, enabling us to suggest the users with certainty higher than a threshold to repeat the audio test on their phones or to take clinical tests if digital diagnosis still fails. 
This study paves the way for a more robust sound-based COVID-19 automated screening system.
\end{abstract}

\noindent\textbf{Index Terms}: COVID-19, sound-based digital diagnosis, ensemble learning, uncertainty estimation 

% \tx{To ask: why only false positive}
% \cm{you say "false predictions" so it's not just false positives... which is fine right?}
% \tx{exactly, that is i mean both false positive and false negative prediction.}
% \tx{change back to false prediction}
\section{Introduction}
%1-2 sentences on covid and traditional testing [done]
%-1-2 sentences on digital testing and sound [done]
%-at the end of this paragraph explain how ground truth is collected re testing: eg crowdsourced from user re have they been tested etc this would avoid problems later when you need to define ‘covid-positives” and negatives
Since the outbreak of the Coronavirus Disease 2019 (COVID-19) in March 2020, over 100 million confirmed cases and 2 million deaths have been identified globally. Frequent and massive screening with targeted interventions is of vital need to mitigate the epidemic~\cite{hunter2020first,atkeson2020economic}.
Currently, the most common screening tool for COVID-19 is the Reverse Transcription Polymerase Chain Reaction (RT-PCR), which is limited by cost, time, and  resources~\cite{Cevik20-Virology,Vogels20-Analytical}. To fight against the virus, researchers' effort has gone into exploring machine learning for fast, contactless, and affordable COVID-19 detection from sounds on smartphones~\cite{schuller2020covid}: COVID-19 is an infectious disease, and most infected people experience mild to moderate respiratory illness~\cite{guan2020clinical}. 
To validate the effectiveness of these approaches, sound data are normally collected with self-reported COVID-19 testing results or more trustworthy COVID-19 clinical testing codes.

% Note that, in general, data recorded from people's device are gathered together with details of their COVID-19 test result either with more trustworthy COVID-19 test code, or trusting the users are telling the truth about their testing status.\jh{not sure the use of the last sentence here? maybe delete from here?} \tx{Cecilia suggested to talk about the data in a general view.}\tx{How about: To validate the effectiveness of the proposed approaches, sound data has been collected with self-reported COVID-19 testing results or more trustworthy COVID-19 testing codes. }

%-list your two problems imbalance and uncertainty [done]
Although recently great progress have been witnessed on cough~\cite{wei2020real,imran2020ai4covid,bagad2020cough,brown2020exploring,laguarta2020covid} and speech-based~\cite{han2020early,pinkas2020sars,asiaee2020voice} COVID-19 detection, there are still unsolved issues which hinder the rollout of this technology to the masses. First, the collected sound data are generally imbalanced, with a small proportion of COVID-19 infected or tested positive participants~\cite{laguarta2020covid,han2020early,imran2020ai4covid}. Such imbalance in training makes the classifier biased to the majority class for a relatively small loss, but it does not mean distinguishable COVID-19 features can be learned from human sounds~\cite{guo2008class,schubach2017imbalance}. This issue is even more detrimental in deep learning as the data are limited and thus balancing solutions are insufficient~\cite{japkowicz2002class}. In addition, even if machine learning can achieve high precision, difficult diagnosis cases (e.\,g., out of training data distribution, noise, sound distortion) are unavoidable~\cite{leibig2017leveraging,qendro2021benefit}. In this respect, information about the reliability of the automated COVID-19 screening is a key requirement to be integrated into diagnostic systems for better risk management. Yet, none of the existing works on sound-based COVID-19 detection takes into consideration the uncertainty in the machine learning prediction. 

In this paper, we propose an ensemble learning-based framework to tackle the training data imbalance and uncertainty estimation challenges, simultaneously. Briefly, when training deep learning models for COVID-19 detection, a number of balanced training sets are generated from the imbalanced data to learn multiple ensemble units. During inference, decisions are fused to maximize data utilization and improve generalization ability. More importantly, we make use of the disagreement across the learned deep learning models as a measure of uncertainty. Softmax probability may indicate the confidence of the prediction, but it tends to overestimate confidence and requires further calibration~\cite{guo2017calibration}. Instead, disagreement from deep ensembles as the uncertainty estimation was proven to better represent the overall model confidence~\cite{lakshminarayanan2017simple}.
 With uncertainty, predictions with low confidence can be identified and these samples could be excluded from digital screening.  The users who produced these samples could be asked to repeat smartphone testing or referred on for different types of testing. As a consequence, this method improves the system performance and patient safety at the same time~\cite{laves2019quantifying}. 
 
To help with this research, we launched a mobile app in April 2020 to crowdsource sound data including breathing, cough, and speech with self-reported COVID-19 testing results\footnote{www.covid-19-sounds.org}. 
In conclusion, our contributions in this paper are summarised as follows,
\begin{itemize}
    \item To handle the limited and imbalanced training data problem, we propose an deep ensemble learning-based framework for COVID-19 sounds analysis, yielding higher AUC and sensitivity compared to other balancing approaches.
    \item To the best of our knowledge, we are the first to investigate the uncertainty of deep learning for sound-based COVID-19 detection, leading to a more reliable and robust automated diagnosis system.
    \item We perform experiments on our collected data with 469 tested positive and 1\,526 healthy control samples, achieving an AUC of 0.74 against 0.62 from an SVM baseline. With the estimated uncertainty, the AUC is further improved up to 0.85.
\end{itemize}

\begin{figure*}[!t]
    \centering
    \includegraphics[width=0.9\textwidth]{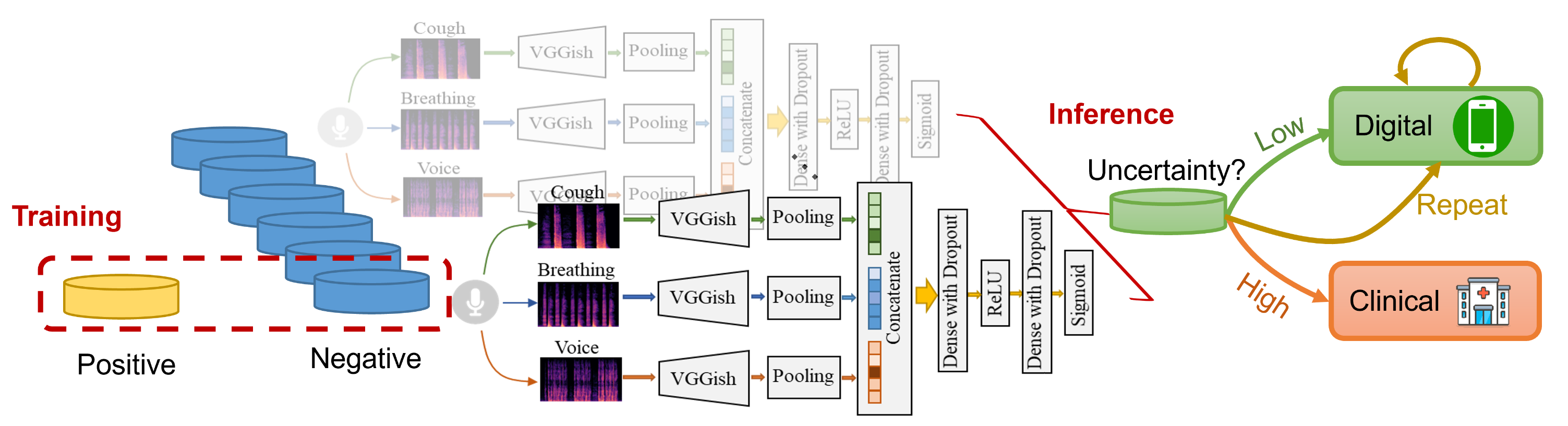}
    \caption{Uncertainty-aware deep ensemble learning-based COVID-19 detection framework. Balanced training sets are generated to train multiple CNN-based models, and probabilities for a testing sample are fused to form the final decision. Simultaneously, the disagreement level across these models as a measure of uncertainty is obtained and used to indicate the reliability of digital diagnose. }
    \label{fig:overview}
\end{figure*}

\section{Related Works}
In~\cite{wei2020real,imran2020ai4covid,brown2020exploring,laguarta2020covid,pahar2020covid}, respiratory sounds, especially coughs, have been analysed and detectable features have been discovered from spectrograms to distinguish COVID-19 coughs from healthy or other disease coughs. At the same time, researchers have exploited speech signals~\cite{han2020early,pinkas2020sars,asiaee2020voice,schullerinterspeech}, which are more natural and informative.
All those studies demonstrate the promise of detecting COVID-19 form sounds, in a non-invasive, inexpensive, and largely available manner.
However, all the above works have suffered from imbalanced data problem. For instance, 92 COVID-19 tested positive and 1079 healthy subjects were included in~\cite{imran2020ai4covid,sharma2020coswara}. %\tx{This is the Coswara data}. 
The imbalance was either dealt with down-sampling, up-sampling or generally not yet tackled in the research~\cite{brown2020exploring,han2021exploring, pahar2020covid, laguarta2020covid,schuller2020detecting}.
However, down-sampling, by discarding samples from the majority class, results in an even smaller size of training data and might lose important discriminative information, while up-sampling, by repeating samples from minority class, might change the original distribution of the data and lead to model overfitting. Moreover, although Synthetic Minority Oversampling Technique (SMOTE) is a more advanced up-sampling method~\cite{chawla2002smote}, it is inherently dangerous as it violates the independence assumption and blindly generalizes the minority area without regard to the majority class.

%The imbalance in previous work was either  ignored~\cite{laguarta2020covid,schuller2020detecting} or simple down-sampling~\cite{ brown2020exploring} and up-sampling~\cite{han2021exploring, pahar2020covid} approaches was adopted to maintain the balance. 
% Up or down-sampling has proven ineffective as these COVID-19 datasets are relatively smalls and...\cm{please explains why they do not work on small data} \tx{Down-sample wastes data but upsample, SMOTE, need @Jing, use repeated samples?}\td{down-sampling discarding samples from majority class results in even smaller size of training data and might lose important discriminative information while up-sampling repeating samples from minority class might change the original distribution of the data and lead to model overfitting. Though Synthetic Minority Oversampling Technique (SMOTE) is a more advanced up-sampling method, it is inherently dangerous since it violates the independence assumption and blindly generalizes the minority area without regard to the majority class. }

There has been a consensus in the literature that uncertainty estimation could be used to aid automated clinical diagnosis, for example for clinical imaging analysis~\cite{leibig2017leveraging}.
%Moreover, none of the existing sound-based COVID-19 screening works considered the uncertainty of the predictions, while (introduction mentioned)
In terms of COVID-19 diagnosis, one work has obtained uncertainty from CT (computerized tomography)  scans to achieve interpretable COVID-19 detection~\cite{ghoshal2020estimating}.
%Another study proposed to exclude some testing samples if the decisions from three independent models are not consistent~\cite{imran2020ai4covid}, but it failed to estimate uncertainty specifically and explicitly, neither conducting further insightful analysis. 
Though softmax probability may indicate the confidence of the prediction, to some extent, it only captures the relative probability that an instance is from a particular class, compared to the other classes, instead of the overall model confidence. Furthermore, it tends to overestimate confidence and thus requires further calibration~\cite{guo2017calibration}.
In general, Bayesian Convolutional Neural Network~\cite{teye2018bayesian} and Monte Carlo Dropout (MCDrop)~\cite{gal2016dropout} \& Monte Carlo Dropweights~\cite{ghoshal2020estimating} have been exploited to estimate uncertainty. However, Bayesian estimation is computationally expensive, which is not an optimal solution for our task with limited training data, while Dropweights in~\cite{ghoshal2020estimating}, keeps dropout on during inference, reducing the model capacity and may leading to lower accuracy~\cite{lakshminarayanan2017simple}. With evidence suggesting that uncertainty from deep ensembles outperforms MCDrop~\cite{lakshminarayanan2017simple, ovadia2019can,wu2020ensemble}, this paper proposed the ensemble learning framework which tackles the data imbalance and uncertainty estimation simultaneously within a unified framework.

\section{Methodology}

In this paper, we focus on developing an uncertainty-aware audio-based covid-19 prediction model. In particular, the basic unit integrates information from three different sound types, i.\,e.,  breathing, cough, and speech. Then, ensemble learning is exploited to handle the highly unbalanced data. Furthermore, an uncertainty estimation can be obtained from the ensemble framework. The proposed framework is illustrated in~Figure~\ref{fig:overview}.

% Our proposed uncertainty-aware COVID-19 detection from sounds framework is illustrated in Figure~\ref{fig:overview}. It exploits ensemble learning and inference to deal with imbalanced data and to capture the uncertainty of the prediction.
% To tackle the challenge caused by small-scale and imbalanced data, during training, we sample from the majority class (healthy controls) to generate balanced training sets, from which we train multiple deep learning models. 
% \jh{deep learning can be deltected? it also applies with SVM models.}
% %This allows to train on balanced classes and taking advantage of the data variability in the majority class.
% During inference, for each unseen testing sample, we perform a forward pass on each model composing the ensemble and fuse the results for the final decision. To obtain the predictive uncertainty, we compute the disagreement level within the model suite, which can reveal the confidence of the digital screening. 
% %Predictive uncertainty provides a better interpretation of the inference by capturing the noise in the observations and the uncertainty in the model parameters. Uncertainty measurements are a robust and efficient trigger on deciding when a human-in-the-loop is required. 

\subsection{COVID-19 Detection Model}\label{sec:model}
Three different modalities are adopted to develop the deep learning-based COVID-19 detection model - the basic unit of ensemble learning. Following previous work~\cite{brown2020exploring}, the CNN model named Vggish~\cite{hershey2017cnn} is applied and adapted as the feature extractor, which is trained on a large-scale audio dataset for audio event classification.
In particular, Mel-spectrums are extracted from each modality and fed into Vggish, which yield a 128-dimensional embedding for each 0.96-second audio segment. It is followed with average pooling along the time-axis to obtain a fixed-sized feature vector from length-varying inputs. The feature vectors of three modalities are then concatenated as the combined features. Finally, another two dense layers with a softmax function are utilized as a binary classifier, the outputs of which can be interpreted as the probability of COVID-19 infection.

\subsection{Ensemble Training and Inference}
Many machine learning approaches struggle to deal with real-world health data because it is common to have imbalanced datasets where the healthy users are a significant majority of the whole set. This is also the case for COVID-19 sound-based detection, where the users who tested positive are the minority class. To tackle this problem, as described in Figure~\ref{fig:overview}, we generate a series of training bags $\{(X_1, Y_1),(X_2, Y_2),...,(X_N, Y_N)\}$ with an equal number of positive and healthy users by re-sampling from the majority healthy class, where $X_i$ denotes the input audio samples within the $i$-th bag and $Y_i$ are the corresponding labels. Note that since some healthy users can be re-used, we can generate numerous bags. Consequently, based on a fixed and shared validation set, we train $N$ neural networks, as introduced in Sec.~\ref{sec:model}, via cross-entropy loss, independently.  During inference, we pass the testing sample $x$ into the ensemble suite and obtain probability-based fusion~\cite{schuller2018interspeech} as the final output, formulated as follows,

\begin{equation}\label{equ:1}
\small
\vspace{-0.04in}
    p(y) = \frac{1}{N}\sum_{i=1}^{N}P(y=1|x,X_i,Y_i,\mathbf{\theta}_i),
\end{equation}
where $P(y|x,X_i,Y_i,\mathbf{\theta}_i)$ is the predicted softmax probability of $i$-th model. If $p(y)$ is higher than 0.5, the given testing sample will be predicted as COVID-19 positive.

\subsection{Uncertainty Estimation}
No matter how good the deep learning model is, difficult cases to diagnose are unavoidable: this could be due to many reasons, including very noisy samples. This highlights
the importance of uncertainty estimation for the digital screening. Considering the incapability of softmax probability to capture model confidence, we define the disagreement level across models within the ensemble suite as the measure of uncertainty. Uncertainty from deep ensembles has also been shown to be more effective than other estimation approaches~\cite{lakshminarayanan2017simple}. To be specific, we use the standard deviation across the $N$ models as the measurement of uncertainty as follows,
\vspace{-0.06in}
\begin{equation}
\small
    \sigma(y) = \sqrt{\frac{1}{N}\sum_{i=1}^{N}(P(y=1|x,X_i,Y_i,\theta_i)-\mu)^2},
    \vspace{-0.06in}
\end{equation}
where $\mu$ is the averaged probability, as $p(y)$ in Eq.~(\ref{equ:1}).

If the uncertainty $\sigma(y)$ is higher than a predefined threshold, it implies that the model is confident enough with its prediction during digital pre-screening. % This might be caused by a noising input or because the audio recording is out of the distribution of our training data. 
Under this circumstance, the system can first request a second or even more repeated audio testing on smartphones. If the uncertainty is still high,  this particular sample could be then referred for further clinical or another testing. 
As a consequence, both system performance and patient safety can be improved.

\section{Evaluation}

\subsection{Dataset}
Given the great potential of audio-based COVID-19 digital screening, we launched an app, namely \textit{COVID-19 Sounds App}, to crowdsource data for research. In the initial registration, users' basic demographic and medical history information are required. Then, users are guided to record and submit breathing, coughing, and short speech audios, together with the latest coronavirus testing results and associated symptoms, if any, every other day. To be more specific, audios are three to five inhalation-exhalation sounds, three voluntary cough sounds, and the participant reading a standard sentence from the screen three times.
%As of February 26$^{th}$ 2021, 68\,324 data samples from 36\,675 users have been collected.

%select criteria, statistics
In this study, we focus on the group consisting of users who declared to have tested positive and ones who declared they have tested negative and declared no symptoms: we call these users ``healthy". To avoid any language confounders in the voice recordings, only English speakers were retained. After a manual quality check,  330 positive users with 469 samples and 919 healthy users with 2\,021 samples were selected. Overall, 58\% of the users are male and more than 60\% are aged between 20 to 49. Demographic and medical history distributions are similar in the two classes. As for pre-processing, we resample all the recordings to 16\,kHz mono audios, and then remove the silence period at the beginning and the end of the recording as in~\cite{han2021exploring}. Finally, audio normalisation by calibrating the peak amplitude to 1 is applied to eliminate the discrepancy across recording devices.

\begin{table}[t]
\caption{Basic statistics of COVID-19 sound data.}
\vspace{-0.10in}
\centering
\resizebox{0.4\textwidth}{!}
    {
    \begin{tabular}{@{}lcccc@{}}
    \toprule
     \multirow{2}{*}{}  & \multicolumn{2}{c}{\textbf{Positive}}    & \multicolumn{2}{c}{\textbf{Healthy}}  \\ \cmidrule(r){2-3} \cmidrule(r){4-5} 
                      & \emph{\#Users}  & \emph{\#Samples} & \emph{\#Users}  & \emph{\#Samples} \\ \midrule
        Train      & 231 & 327                 & 820 &1871                \\
        Validation & 33  & 44                  & 33  &56                \\
        Test       & 66  & 98                  & 66  &94                  \\            
     \bottomrule
    \end{tabular}}
\label{tab:dataset}
\vspace{-0.15in}
\end{table}

\subsection{Experimental setup}
To evaluate the proposed framework, for the positive group, we hold out 10\% and 20\% of users as validation and testing sets, and then use the remaining data for training. Correspondingly, we select the same of healthy users for validation and testing (see Table~\ref{tab:dataset}). Furthermore, to generate a balanced training set for each ensemble unit $(X_i, Y_i)$, 231 users are randomly selected from 820 negative tested users. The number of hidden units of dense layers in our model is set to 64 and 2, respectively. When training, our batch size is 1, the learning rate is 0.0001 with a decay factor of 0.99 and we use cross-entropy loss and Adam optimiser. Early stopping is applied on the validation set to avoid over-fitting.
All experiments are implemented in Python and TensorFlow. Feature extractor is initialled by pre-trained VGGish model and then updated with the following dense layers jointly.

Moreover, baselines from the latest literature are conducted for performance comparison. 
In addition to deep models, acoustic feature-driven SVM achieved state-of-the-art performance in sound-based COVID-19 detection due to its effectiveness and robustness in small data learning~\cite{brown2020exploring,han2020early,han2021exploring}. Therefore, we repeat the experiments in~\cite{han2021exploring} by using the openSMILE toolkit to extract 384 acoustic features and SVM with linear kernel and complex constant $C=0.01$ as the classifier. For both SVM and deep models, to test the superior performance of ensemble learning, we compare training one single model with all samples, with balanced samples by down or up-sampling, against training $N=10$ models in an ensemble learning manner. For down-sampling balancing, we randomly discard some healthy users, while for up-sampling, synthetic minority oversampling techniques (SMOTE)~\cite{chawla2002smote} is performed to generate synthetic observations of the minority class. This is the most commonly adopted technique for imbalanced data.

To justify the performance of the proposed framework for COVID-19 screening and diagnosis, we calculate the following metrics: 
\textbf{\textit{Sensitivity}}, also named true positive rate or recall defined by $TP/(TP+FN)$, 
\textbf{\textit{Specificity}}, also referred to as true negative rate formulated by $TN/(TN+FP)$,
and \textbf{\textit{ROC-AUC}}, the area under receiver operating characteristics curve. which measures the overall sensitivity and specificity at various probability thresholds.
Besides, for both the baseline and our proposed methods, the mean and standard deviation of the aforementioned metrics across 10 runs are reported.

\subsection{Results}
\subsubsection{Performance of Ensemble Learning}
The overall comparison is presented in Table~\ref{tab:results}. 
First, deep learning is more vulnerable than the SVM model with imbalanced training data. When the CNN model achieves an ROC-AUC of 0.69, against 0.60 of SVM, the sensitivity is very low because the model is biased to the healthy class, and the very high specificity of 0.98 is practically meaningless. 
Second, resampling can improve the performance, especially in terms of sensitivity for both SVM and deep learning, since a balanced training set is guaranteed.  Compared to down-sampling, up-sampling performs better, as expected, because more data samples are available for parameter learning.
Third, ensembles can boost the performance of both SVM and CNN.
Last but not least, our CNN ensemble framework outperforms all the baselines and achieves an AUC of 0.74 with a sensitivity and a specificity close to 0.7, demonstrating the superiority of exploiting ensemble learning and deep CNN network for COVID-19 detection from imbalanced data. In addition, the commonly used majority voting fusion method~\cite{schuller2018interspeech} was also validated in comparison to our proposed probability-based fusion: an AUC of 0.74 with a sensitivity of 0.62 and a specificity of 0.70 was obtained. Given the same AUC but lower sensitivity, we confirm that probability-based fusion is more promising.

\begin{figure}[t]
    \centering
    \includegraphics[width=0.34\textwidth]{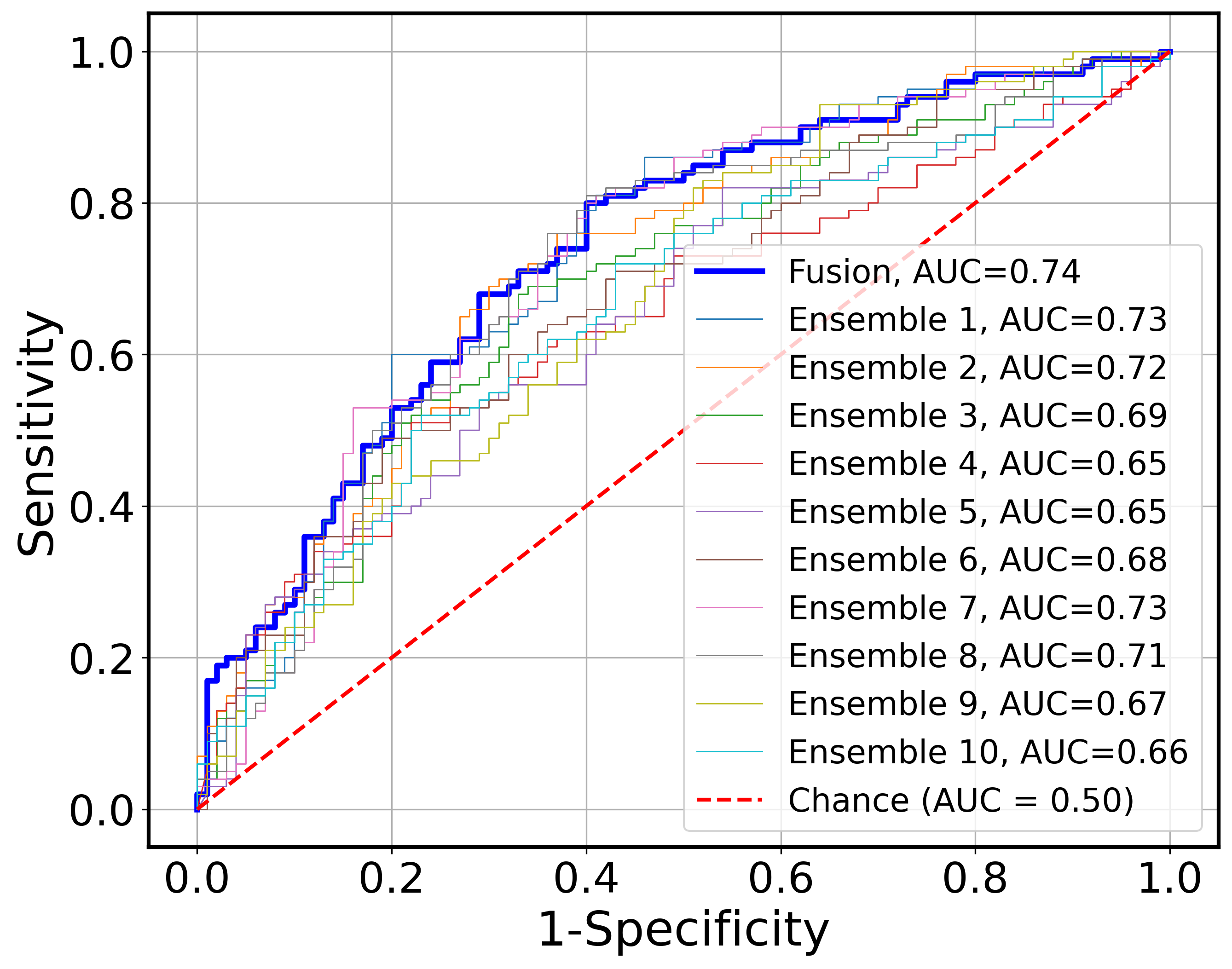}
    \vspace{-0.10in}
    \caption{ROC curves of COVID-19 detection, where the curve of each ensemble unit model is shown separately. The ROC curve after probability-based fusion is shown in bold.}
    \label{fig:aucs}
    \vspace{-0.25in}
\end{figure}

To further inspect the ensemble suite, we visualise the ROC curves for each unit model in Figure~\ref{fig:aucs} for comparison. All ROC curves are above chance level but the model variance is not negligible. A plausible explanation is that we only have a small training dataset for each unit. However, after probability-based fusion, the ROC curve is generally higher than the other curves.  All these demonstrate that ensembles can improve the robustness and generation ability of machine learning models.

To conclude, by integrating multiple networks trained from different balanced data, our proposed approach achieves state-of-the-art with the AUC of 0.74, the sensitivity of 0.68 and the specificity of 0.69 for COVID-19 detection.
\begin{table}[t]
\caption{Performance comparison with Mean(Std) for ROC-AUC, Sensitivity, and Specificity reported for Single model (SM) and Ensemble model.}
\vspace{-0.10in}
\centering
\resizebox{0.45\textwidth}{!}{%
\begin{tabular}{lcccc}
\toprule
\multicolumn{2}{c}{}                              & ROC-AUC        & Sensitivity         & Specificity         \\ \midrule
\multirow{2}{*}{SM imbalanced data} & SVM & 0.60       & 0.54       & 0.57       \\
                                                  & CNN & 0.69       & 0.15       & 0.98       \\\hline
\multirow{2}{*}{SM down-sampling} & SVM & 0.60(0.03) & 0.68(0.05) & 0.45(0.05) \\
                                                  & CNN & 0.68(0.04) & 0.62(0.04) & 0.63(0.06) \\\hline
\multirow{2}{*}{SM up-sampling}   & SVM & 0.62(0.02) & 0.53(0.02) & 0.63(0.02) \\
                                                  & CNN & 0.70(0.04) & 0.52(0.02) & 0.77(0.05) \\\hline
\multirow{2}{*}{\textbf{Ensemble model}}                   & SVM & 0.66(0.04) & 0.63(0.05) & 0.62(0.04) \\
                                                  & \textbf{CNN} & \textbf{0.74(0.03)} & \textbf{0.68(0.05)} & 0.69(0.06) \\
\bottomrule
\end{tabular}}
\label{tab:results}
\vspace{-0.1in}
\end{table}

\begin{figure}[t]
  \centering
  \subfigure[Uncertainty distrubution.]{
  \label{fig:dis}
  \includegraphics[width=0.28\textwidth]{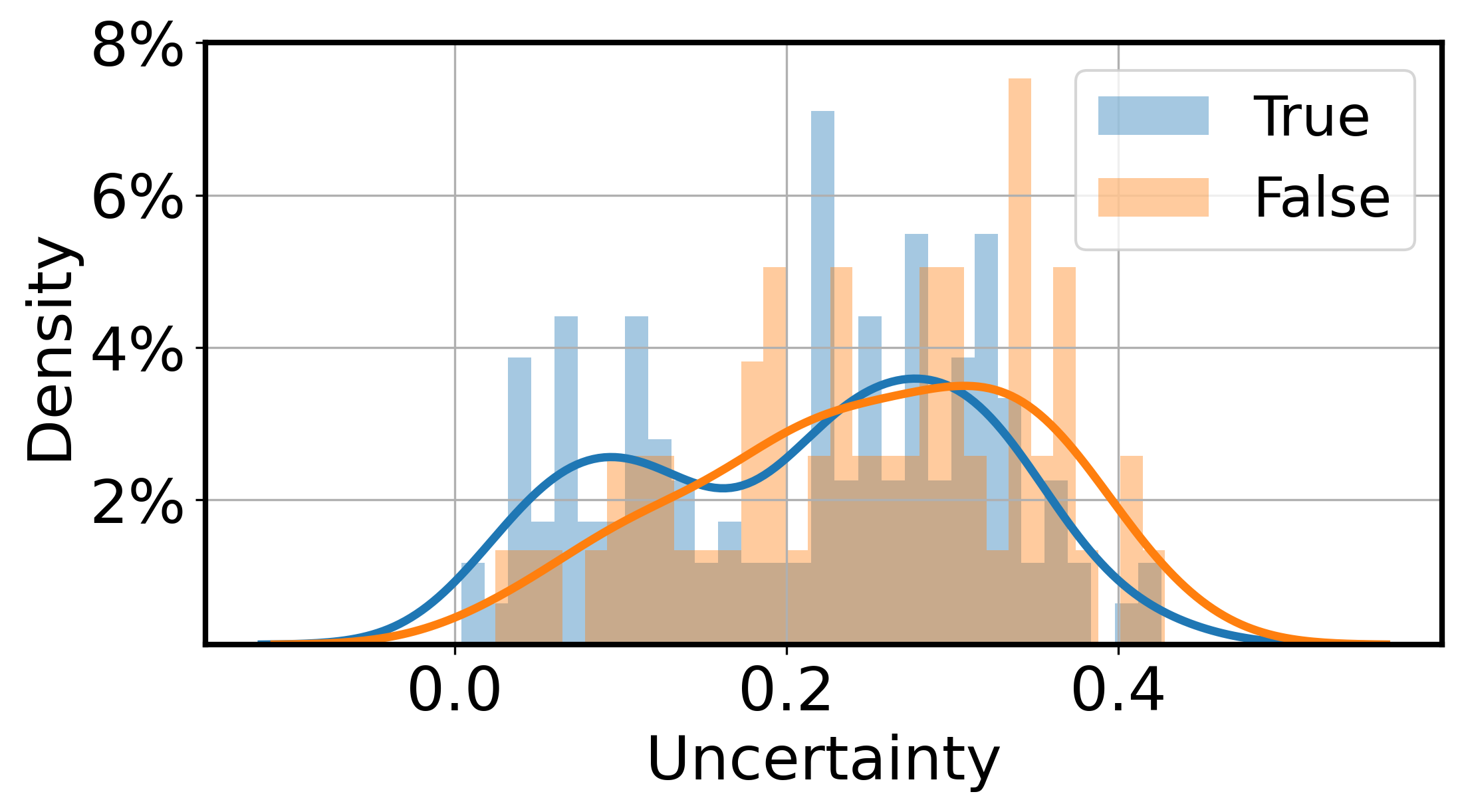}}
 
  \subfigure[Uncertainty threshold.]{
  \label{fig:auc_th}
  \includegraphics[width=0.2\textwidth]{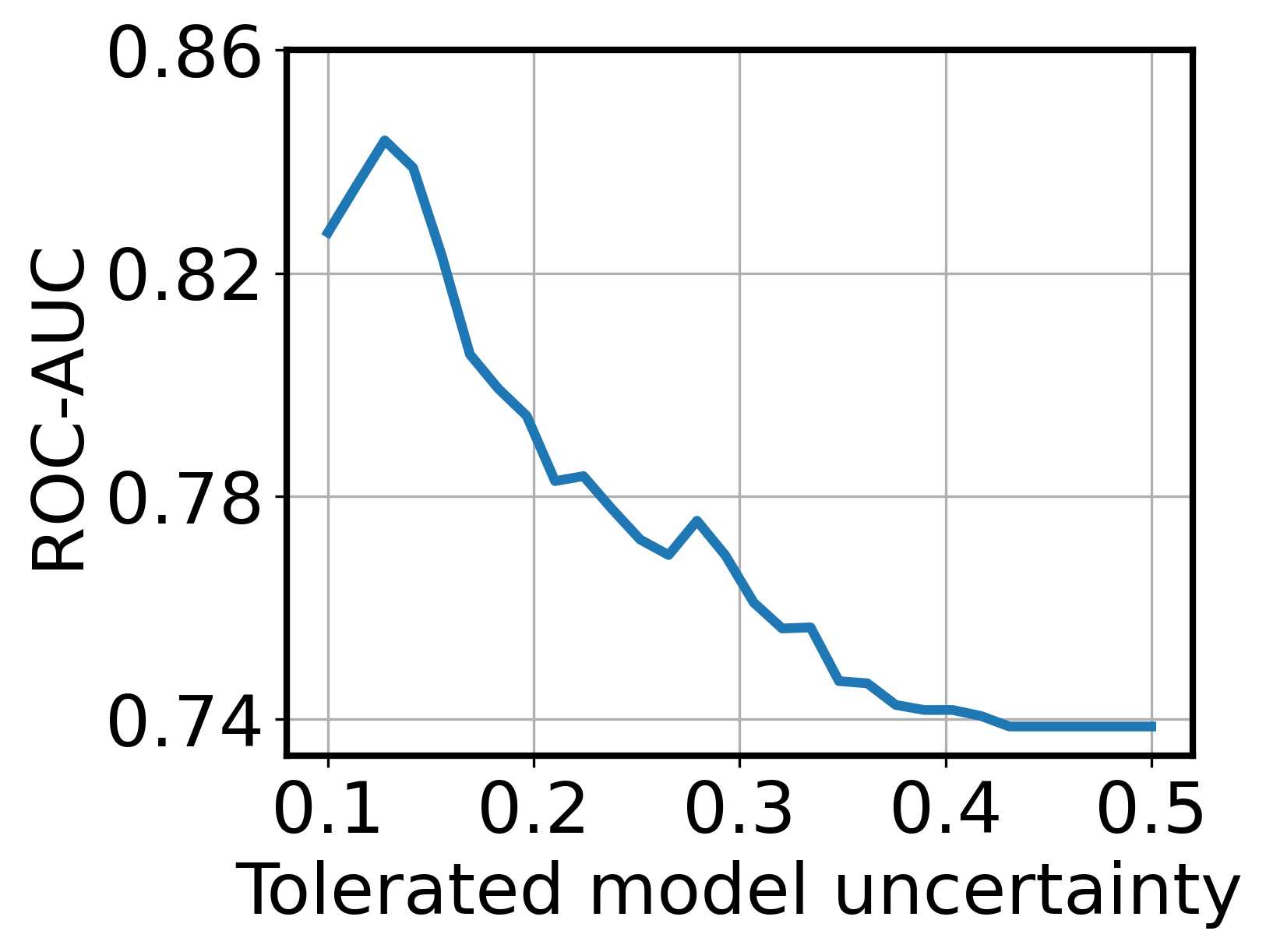}}
 \subfigure[Remained fraction.]{
  \label{fig:auc_po}
  \includegraphics[width=0.2\textwidth]{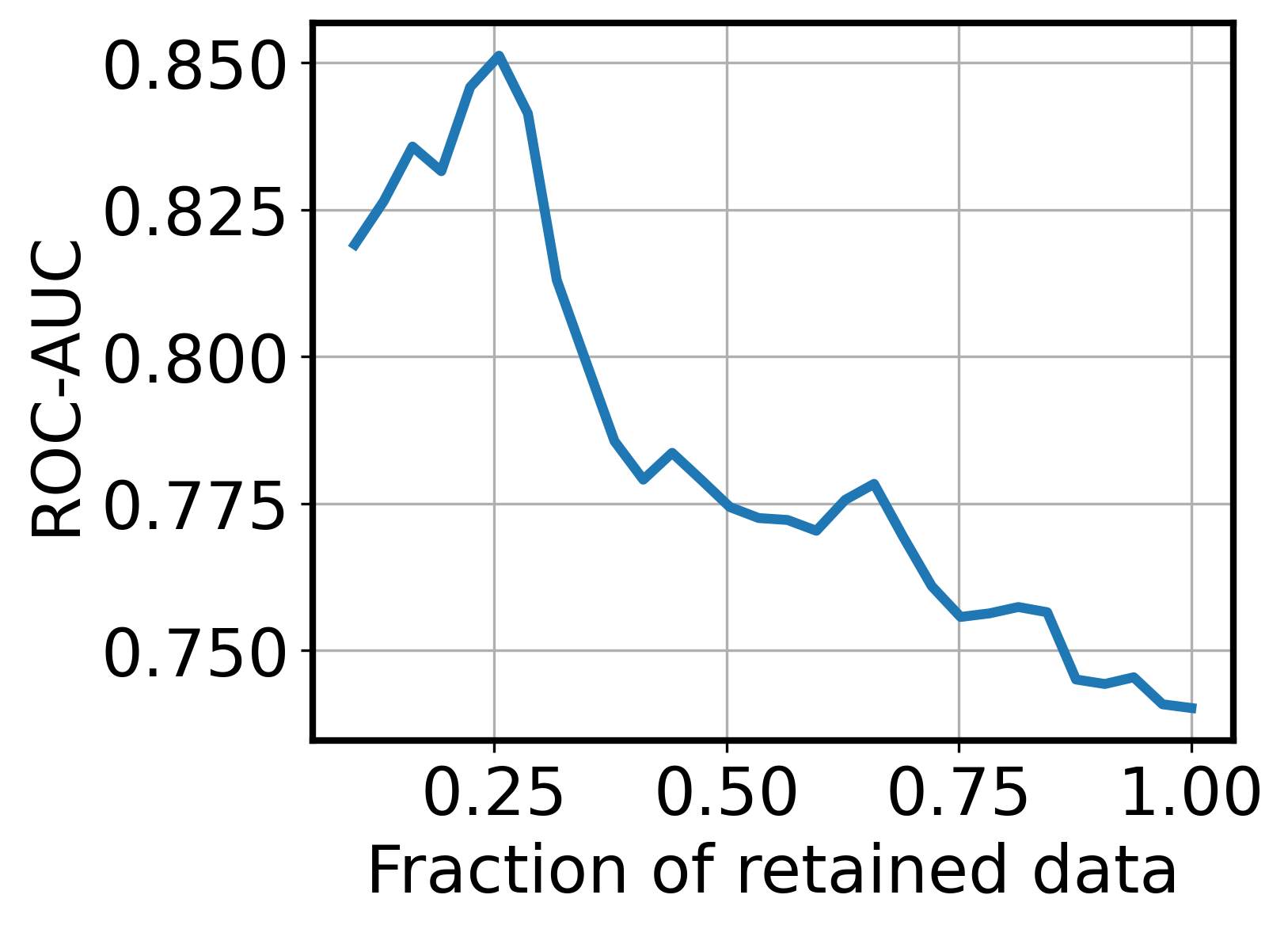}}
  \vspace{-0.15in}
  \caption{Performance for uncertainty-aware referral.}
  \vspace{-0.25in}
\end{figure}
\subsubsection{Estimation and Application of Uncertainty}
By using the standard deviation  across ensemble models as the measure of uncertainty, we are able to understand the confidence of the predictions. We illustrate the uncertainty distribution in Figure~\ref{fig:dis}. The density of high uncertainty values for false prediction is significantly higher than that of true prediction, indicating that our approach succeeds in identifying less confident predictions when making an incorrect diagnosis.

Inspired by the findings, we set up thresholds to exclude some testing cases when the uncertainty is higher than a given value. Results from our dataset are shown in Figure~\ref{fig:auc_th}, from which we can find that when keeping only testing samples with an uncertainty lower than 0.2, the AUC can be improved from 0.74 to 0.79, and when the threshold is 0.12, the highest AUC value is reached at 0.85. Note that the drop on the left side of the curve is caused by a small number of samples to calculate the metric. These results show that studying predictive uncertainty and finding the optimal threshold can significantly improve automatic digital prediction.

Referral for further testing can impose a range of costs which can vary from a cost-negligible ``repeat the audio test on your phone" to a clinician inspection or visit. 
In this respect, we inspect the AUC on different factions of retained data with the uncertainty lower than a certain threshold (assuming the samples above this threshold would be referred on). Figure~\ref{fig:auc_po} shows that the AUC climbs from 0.74 to 0.76 when 15\% of the samples with the highest uncertainty are excluded.  When only remaining 25\% data with the lowest uncertainty, the AUC can increase to 0.85. This indicates that our measurement for uncertainty is informative and helpful for a more robust automatic COVID-19 diagnosis system.
%\cm{I think part of this needs to go back to the intro..ie whne we say that we can refer some samples clinically and therefore improve accuracy we can say that this is a paramater we can control but that with just x percent of samples referred we can increase the accuracy of Y, for example}\cm{also 25 percent seems a bit high..can you check other lower percentages?}

\section{Conclusions}
In this work, a sound-based machine learning approach has been proposed to discriminate COVID-19 cases from health controls. Ensemble learning has been explored to solve the imbalanced training data challenge, yielding favorable performance gain on an in-the-wild crowdsourced dataset. In addition, uncertainty has been measured via the disagreement across ensembles, and thus enables confidence-informed digital diagnose. To the best of our knowledge, we are the first to introduce uncertain-aware deep learning approach to sound-based COVID-19 detection studies.  

For future work, we plan to gain deeper understanding on the estimated uncertainty, exploring  how each modality (breathing, cough, speech) contributes to the overall uncertainty. Our data collection is still ongoing which will yield more data to train our framework for further performance evaluation.

%1) the epistemic and aleatoric uncertainty individually~\cite{ghoshal2019estimating}, 2) how each modality (coughing, breathing, voice) contributes to the overall uncertainty
%, and 3) how to incorporate uncertainty in gradient back propagation to further improve the performance. Besides, 
% Our data collection is still ongoing which will yield more data to train and valuate our framework for further performance testing.\cm{don't give away good ideas...the aleatoric thing keep to yourself}

\section{Acknowledgements}
This work was supported by ERC Project 833296 (EAR). 
%.\cm{do add the Huawei donation!}
% \cm{Tong have you asked the office how they would like you to acknowledge your funding? there is usually a precise way...} .%and the UK Cystic Fibrosis Trust.
% %We further thank everyone who volunteered their data. \cm{we have already thanked them in the papers describing the app and the data collection}
% %Tong Xia is funded by Huawei PhD scholarship.
% \cm{lorena, we won't thank Nokia here as they have an issue being on the same paper as Huawei so we simply can't. I think it's ok as this won't be part of your PhD..}
% \jh{it reminds me another issue. I would suggest not to mention huawei donation during submission, just in case some reviewers dislike huawei... The acknowledge part can be revised after acceptance at which stage we can add it back.}
% \lo{agree with both Cecilia and Jing ;) }
% \tx{OK, will remove huawei}
\clearpage
\balance
\bibliographystyle{IEEEtran}
\bibliography{mybib}
\end{document}